\documentclass[aps,prb,longbibliography,showpacs,reprint,superscriptaddress]{revtex4-1}

\usepackage{graphics}
\usepackage{graphicx}
\usepackage[fleqn]{amsmath}
\begin{document}

\title{Traveling Wave Parametric Amplifier based on a chain of Coupled Asymmetric SQUIDs}

\author{M. T. Bell}
\affiliation{Engineering Department, University of Massachusetts Boston, Boston, Massachusetts 02125}
\affiliation{Department of Physics and Astronomy, Rutgers University, Piscataway, New Jersey 08854}

 \author{ A. Samolov}
\affiliation{Engineering Department, University of Massachusetts Boston, Boston, Massachusetts 02125}

\date{\today}

\begin{abstract}
A traveling wave parametric amplifier (TWPA) composed of a transmission line made up of a chain of coupled asymmetric superconducting quantum interference devices (SQUIDs) is proposed. The unique nature of this transmission line is that its nonlinearity can be tuned with an external magnetic flux and can even change sign. This feature of the transmission line can be used to perform phase matching in a degenerate four-wave mixing process which can be utilized for parametric amplification of a weak signal in the presence of a strong pump. Numerical simulations of the TWPA design have shown that with tuning, phase matching can be achieved and an exponential gain as a function of the transmission line length can be realized. The flexibility of the proposed design can realize: compact TWPAs with less than 211 unit cells, signal gains greater than 20 dB, 3 dB bandwidth greater than 5.4 GHz, and saturation powers up to -98 dBm. This amplifier design is well suited for multiplexed readout of quantum circuits or astronomical detectors in a compact configuration which can foster on-chip implementations.
\end{abstract}


\maketitle

%

\section{Introduction}
\indent Over the past decade Josephson parametric amplifiers have proven essential in experiments studying quantum jumps \cite{VijayPRL11} , tracking quantum trajectories \cite{HatridgeSci13, MurchNat13, WeberNat14} and real-time monitoring and feed-back control  \cite{HatridgeSci13,VijayNat12,RisteNat13}  of quantum bits (qubits). In parametric amplifiers high gain is achieved when the signal to be amplified interacts with a nonlinear medium for as long as possible. State-of-the-art Josephson parametric amplifiers utilize resonant circuits to increase the interaction time of the signal with the nonlinear medium \cite{HatridgePRB11,BergealNatPhy10,BergealNat10,RochPRL12}. As a consequence the instantaneous bandwidth and maximum input power allowed are significantly reduced. Such limitations have renewed interest in superconducting traveling wave parametric amplifiers (TWPA) \cite{EomNatPhy12,MohebbiIEEETransMW09,YaakobiPRB13,YaakobiPRB13Err,OBrienPRL14,SweenyIEEETransMag85,YurkeAPL96,BockstiegelJLTP14,TCWhiteAPL} which achieve long interaction times with the nonlinear medium by extending the electrical length over a long transmission line instead of multiple bounces in a resonant circuit, as a result TWPAs do not suffer from the same bandwidth and dynamic range limitations that cavity based amplifiers do. The main challenges in TWPA designs is that optimum gain is achieved when phase matching conditions are met. Superconducting TWPAs have been investigated by many groups, thus far taking one of two approaches, either utilizing a transmission line composed of a series array of Josephson junctions \cite{MohebbiIEEETransMW09,YaakobiPRB13,YaakobiPRB13Err,OBrienPRL14,SweenyIEEETransMag85,YurkeAPL96,TCWhiteAPL} or a transmission line utilizing the nonlinear kinetic inductance of a narrow superconducting wire \cite{EomNatPhy12,BockstiegelJLTP14,ChaudhuriArXiv15}. These investigations have revealed the need for engineered dispersion to be introduced into the transmission line to facilitate phase matching \cite{EomNatPhy12,YaakobiPRB13Err,OBrienPRL14,ChaudhuriArXiv13}. Designs which utilize periodic loading  \cite{EomNatPhy12,BockstiegelJLTP14,ChaudhuriArXiv15} and the addition of resonant elements \cite{OBrienPRL14,TCWhiteAPL} to facilitate phase matching have shown promise, however at the expense of increased complexity, higher tolerances, and longer propagation lengths $ (2\text{ cm} - 1\text{ m}) $. 
We propose an alternative approach and utilize the nonlinear properties of a one dimensional chain of coupled asymmetric SQUIDs as a transmission line in a TWPA to achieve phase matching and show that exponential gain can be realized over a wide bandwidth.\\
\indent The proposed TWPA utilizes the tunable nonlinearity of a one dimensional chain of asymmetric SQUIDs with nearest neighbor coupling through mutually shared Josephson junctions as a transmission line to overcome phase matching limitations. A magnetic flux $\Phi$ threads each SQUID to allow for tunability of the linear and nonlinear properties of the transmission line. A weak signal to be amplified and a strong pump tone will be incident on the input port to the transmission line. Due to the nonlinearity of the transmission line the weak signal will undergo parametric amplification through a degenerate four-wave mixing (FWM) process \cite{Agrawal2001,Boyd2008}. The amplification process is the most efficient when the total phase mismatch is close to zero. However due to the nonlinearity of the transmission line a strong pump modifies phase matching through self and cross phase modulation resulting in a phase mismatch. The linear dispersion of the transmission line along with spectral separation of the signal and pump angular frequencies can be used to compensate for the nonlinear phase mismatch. The unique feature of the proposed TWPA is that the linear and nonlinear dispersion can be tuned with $\Phi$, and the nonlinearity can even change sign. By adjusting $\Phi$ for a given pump power, phase matching can be achieved.

\section{TWPA Design}
\begin{figure}[h!]
\includegraphics[width=8.5cm]{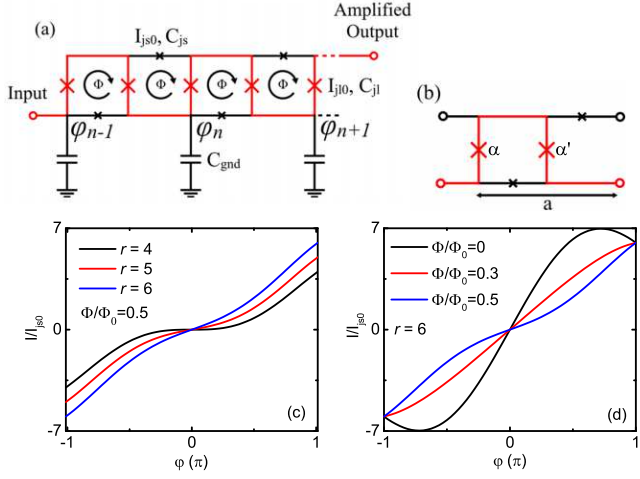}   
\caption{(color online). TWPA based on a chain of asymmetrically coupled SQUIDs (a) Circuit schematic of the proposed TWPA. Each unit cell of the TWPA is threaded with a magnetic flux $\Phi$ and has a parasitic capacitance to ground $C_{gnd}$. The coupled SQUID transmission line can realistically be implemented in a coplanar transmission line geometry. The geometrical inductance of the transmission line can be made negligible in comparison to the Josephson inductance of the coupled asymmetric SQUIDs making up the line.  (b) Unit cell of the TWPA. (c) and (d) Current phase relation of a unit cell for various ratio’s $r$ and $\Phi/\Phi_0$. \label{fig1}}
\end{figure}

\indent The design of the proposed TWPA is shown in Figure \ref{fig1}(a). Each cell of the transmission line is an asymmetric SQUID with a single "small" Josephson junction with critical current $I_{js0}$  and capacitance $C_{js}$ in one arm and two "large" Josephson junctions with critical current $I_{jl0}$ and capacitance $C_{jl}$ in the other arm.
\\ Adjacent cells to one another are coupled via the large Josephson junctions. A feature of this arrangement is that for an even number of asymmetrical SQUIDs in the chain, the Josephson energy $E_j (\phi)$ remains an even function of the phase difference $\phi$ across the chain for any value of $\Phi$ \cite{BellPRL12}. The proposed TWPA requires a long transmission line with many unit cells, this allows to neglect boundary effects and focus on translationally invariant solutions. At arbitrary $\Phi$ the transmission line remains symmetric under the translation by two cells. The defined unit cell of the transmission line is composed of two large and two small junctions (Figure \ref{fig1}(b)). Each unit cell is of length $a$ and has a capacitance to ground of $C_{gnd}$. The "backbone" of the unit cell is made up of "large" Josephson junctions highlighted in red, which are designed to have Josephson energies, $E_{jl0}$, two orders of magnitude larger than the charging energy, $E_{cl}=e^2/(2C_{jl})$, of the junction. In this case quantum fluctuations of the phase across individual junctions are small $\propto \exp[-(8 E_{jl0}/E_{cl} )^{1/2} ]$ \cite{Tinkham1996} and thus a classical description of the system can be used. The phases on the two large junctions for each unit cell are $\alpha$ and $\alpha^{'}$, and the total phase across the unit cell is $\varphi=\alpha+\alpha^{'}$. The Josephson energy $E_{j}(\varphi)$ of the unit cell is minimized when $\alpha=\alpha^{'}$, and a gauge was chosen such that an external magnetic flux would induce phases $2\pi\Phi/\Phi_{0}$ on the small junctions. An expansion of the current phase relation of the current flowing through the backbone of the unit cell around $\varphi=0$ is given by
\begin{align}\label{eq1}
\begin{split}
I(\varphi)  =I_{js0} \Big[\frac{r}{2}+2\cos&\Big(2\pi\frac{\Phi}{\Phi_0}\Big) \Big]\varphi - \\
  I_{js0}& \Big[\frac{r}{48}+\frac{1}{3} \cos\Big(2\pi\frac{\Phi}{\Phi_0}\Big)\Big]\varphi^3, 
\end{split}
\end{align}
\noindent where $\Phi_0= h/2e$ is the flux quantum, $h$ is Planck's constant, $e$ is the electron charge, and $r=I_{jl0}/I_{js0}$. For $\Phi/\Phi_0 = 0.5$ there exists a critical $r_0 = 4$ where the linear term in $I(\varphi)$ goes to zero and the cubic term dominates. At this point the inductance $L\propto\big(dI(\varphi)/d\varphi\big)^{-1}$ of the unit cell is the largest. This is the unique feature of the superinductor \cite{BellPRL12} which is similar by design to the proposed TWPA. As for the proposed TWPA a large inductance of the unit cell is not desired, however tunability of the nonlinearity of $I(\varphi)$ is. Fig. \ref{fig1}(c) and \ref{fig1}(d) show the current phase relation of a unit cell of the TWPA for various $r$ and $\Phi$. Fig. \ref{fig1}(c) and equation (\ref{eq1}) show that the nonlinearity at $I(\varphi\approx0)$  is always positive at full frustration $\Phi/\Phi_0 = 0.5$ for $r < 16$. Fig.\ref{fig1}(d) and equation (\ref{eq1}) show that for certain $r$ values $(\text{example } r = 6)$ by adjusting $\Phi$ the nonlinearity can be tuned over a wide range, and can even change sign from negative to positive. By tuning the nonlinearity it is possible to optimize the parametric amplification efficiency of the FWM process.

Using equation (\ref{eq1}) a nonlinear wave equation was derived to describe the node flux $\varphi_n$ shown in Fig. \ref{fig1}(a) along the length of the transmission line\cite{YaakobiPRB13}. Assuming constant $C_{js}$, $C_{jl}$, $C_{gnd}$, $I_{jl0}$, and $I_{js0}$ along the length of the transmission line, no dissipation, and the continuum approximation for a wave-type excitation $(\lambda\gg a)$ the following wave equation is derived
\begin{align} \label{eq2}
\begin{split}
\frac{a^2}{L}\Big[\frac{r}{2}+2\cos\Big(\frac{2\pi\Phi}{\Phi_0}\Big)\Big]& \frac{\partial^2 \varphi}{\partial z^2}+a^2 C_{js}\Big(\frac{r}{2}+2\Big)\frac{\partial^4\varphi}{\partial t^2 \partial z^2} - \\
  C_{gnd}&\frac{\partial^2\varphi}{\partial t^2} - \gamma \frac{\partial}{\partial z}\Big[\Big(\frac{\partial\phi}{\partial z} \Big)^3 \Big]=0, 
\end{split}
\end{align}
\noindent where $L=\varphi_0/I_{js0}$  and $\varphi_0=\Phi_0/(2\pi)$. The first three terms of the wave equation represents the linear contributions to the dispersion on the transmission line due to the distributed inductances and capacitances and how they can be tuned with $r$ and $\Phi$. The fourth term describes the nonlinearity and how the nonlinear coupling coefficient, $\gamma = \big[a^4/(\varphi_0^2 L)\big]\big[(r/48)+(1/3)\cos\big(2\pi\Phi/\Phi_0\big )\big]$  can be tuned with $\Phi$. The solution to equation (\ref{eq2}) is assumed to be four traveling waves, where in the degenerate case the two pump angular frequencies $\omega_p$ are equal, a signal $\omega_s$, and a generated idler tone $\omega_i=2\omega_p-\omega_s$. By taking the slowly varying envelope and undepleted pump approximations a set of coupled mode equations is derived to describe the propagation of the signal and idler traveling waves:
\begin{align}
\frac{\partial a_s}{\partial z}-\frac{i3\gamma k_p^2 k_i k_s(2k_p-k_i)a_i^{*}|A_{p0}|^2}{8\omega_s^2C_{gnd}}e^{i\kappa z}=0, \label{eq3}\\
\frac{\partial a_i}{\partial z}-\frac{i3\gamma k_p^2 k_s k_i(2k_p-k_s)a_s^{*}|A_{p0}|^2}{8\omega_i^2C_{gnd}}e^{i\kappa z}=0, \label{eq4}
\end{align}
\noindent where $a_s$ and $a_i$ are the complex signal and idler amplitudes, $k_m$ is the wave vectors of the pump, signal, and idler $(m=\{p,s,i\})$ (see Appendix A), $ \Delta k=k_s+k_i-2k_p$ is the phase mismatch due to the linear dispersion in the transmission line, the total phase mismatch including self and cross phase modulation is $\kappa=-\Delta k+2\alpha_p-\alpha_s-\alpha_i$, and $|A_{p0}|$ is the undepleted pump amplitude. The terms $\alpha_m\propto \gamma|A_{p0} |^2$, are the self and cross phase modulation of the wave vectors per unit length. 


Equations (\ref{eq3}) and (\ref{eq4}) are similar to well established fiber parametric amplifier theory \cite{Agrawal2001, Boyd2008} and have the following solution to describe the power gain of the signal in the presence of a strong pump with zero initial idler amplitude $G_s=|\cosh (gz)-i(\kappa/2g)\sinh (gz)|^2$ where the exponential gain factor is 
\begin{align}
g=\sqrt{\bigg(\frac{k_s^2k_i^2(2k_p-k_s)(2k_p-k_i)\omega_p^4}{k_p^6\omega_i^2\omega_s^2}\bigg)\alpha_p^2-\bigg(\frac{\kappa}{2}\bigg)^2}. \label{eq5}
\end{align}
At nearly phase matching conditions $\kappa\approx 0$, $g$ is positive and real and the signal gain has an exponential dependence on the length of the transmission line $G_s=|e^{gz}/2|^2$. For small pump amplitudes $|A_{p0}|\cong |A_{s0}|$ and $\omega_s \cong \omega_p$ phase matching is not a concern. For larger pump amplitudes the parametric amplification process losses phase matching through self-phase modulation of the pump, when this happens $g$ becomes small and $\kappa$ is large, the gain scales quadratically as a function of TWPA length. 

The phase mismatch due to the large pump amplitude can be compensated with the linear dispersion along the transmission line if $\alpha_p$ and $\Delta k$ are of opposite sign,
\begin{align}
\kappa=-\alpha_p \frac{2\omega_p^2}{k_p^3}\bigg(\frac{k_s^3}{\omega_s^2}+\frac{k_i^3}{\omega_i^2}-\frac{k_p^3}{\omega_p^2}\bigg)-\Delta k. \label{eq6}
\end{align}
 \begin{figure}[h!]
 \includegraphics[width=8.5cm]{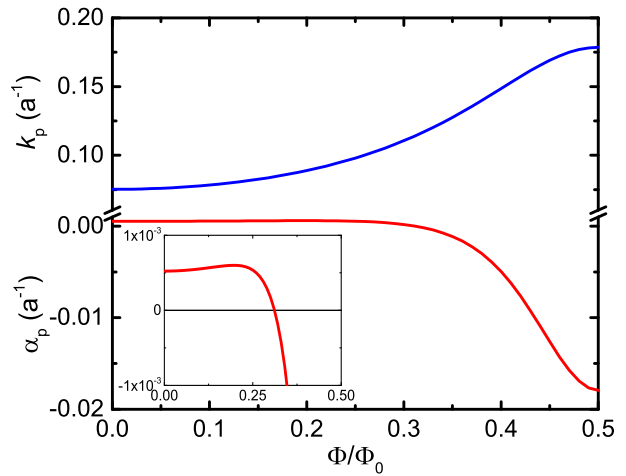}   
 \caption{(color online) Pump tone wave vector and pump self-phase modulation per unit length $a$ as a function of magnetic flux $\Phi/\Phi_0$. The inset shows the change in sign of $\alpha_p$ versus $\Phi/\Phi_0$. \label{fig2}}
 \end{figure}

\section{Numerical Simulations}
From this point on numerical results are presented for a realizable set of parameters for the proposed TWPA. For each unit cell $r = 6$, $C_{gnd} = 50 \text{ fF}$, $C_{js} = 50 \text{ fF}$, $C_{jl} = rC_{js}$, $I_{js0} = 1 \, \mu\text{A}$, and $I_{jl0} = rI_{js0}$. In choosing $C_{gnd}$ and the inductance of the large junctions $ L_{jl}= \Phi_0/(rI_{js0})$ which ultimately sets the characteristic impedance of the transmission line special attention was made to achieve an impedance near $50 \, \Omega$ over the tunable range of the TWPA in order to maintain compatibility with commercial electronics. A realizable unit cell size based on our fabrication process is $a = 8 \, \mu\text{m}$ \cite{BellPRL12,BellPRL14}. The traveling waves used in the numerical results are a pump tone with angular frequency $\omega_p/(2\pi) = 6.5 \text{ GHz}$ and power $-76 \text{ dBm}$ which is equivalent to $I_{prms}\approx 0.6 \, \mu\text{A}$ , the signal angular frequency $\omega_s$ was varied in most cases and the idler angular frequency is $\omega_i=2\omega_p-\omega_s$ with initial signal and idler power levels $80 \text{ dB} \text{ and } 160 \text{ dB}$ lower than the pump power respectively.  

Shown in Figure \ref{fig2} is the dependence of $k_p$ and $\alpha_p$ on $\Phi/\Phi_0$. From the inset in Figure \ref{fig2} it can be seen that $\gamma$ and as a result $\alpha_p$ changes sign from positive to negative for $\Phi>0.31\Phi_0$ and more importantly is of opposite sign to $\Delta k\geq 0$ (see Appendix A) for this transmission line. By adjusting $\Phi$ and at a particular $\omega_p$ and $\omega_s$ it is possible to utilize $\Delta k$ which increases with $\Phi$ in the transmission line to compensate the phase mismatch due to self-phase modulation of the pump.
\begin{figure}[h!]
\includegraphics[width=8.5cm]{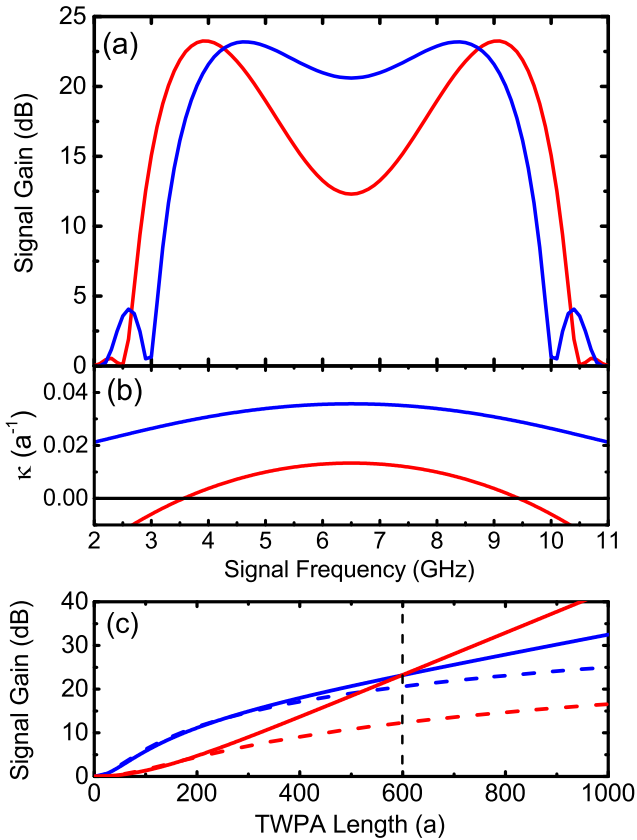}   
\caption{(color online) Calculated gain of the proposed TWPA. In all three panels the color red and blue represent flux tunings (pump powers) of $\Phi/\Phi_0 =0.45 \, (-76 \text{ dBm})$ and $\Phi/\Phi_0=0.5 \, (-73 \text{ dBm})$ respectively.  (a) Signal gain in dB as a function of signal frequency. (b) Phase mismatch as a function of signal frequency. (c) Dependence of the signal gain as a function of transmission line length in units $a$. Solid red and blue lines correspond to $\omega_s/(2\pi)=9.1 \text{ GHz} \text{ and } 8.4 \text{ GHz}$ respectively. Dashed red and blue lines correspond to  $\omega_s/(2\pi)=6.5 \text{ GHz}$. As can be seen the region of exponential gain is $\kappa\approx0$. According to equation (\ref{eq5}) the most optimal gain does not necessarily occur at perfect phase matching $\kappa=0$ due to the pre-factor to $\alpha_p^2$. At significant $\kappa$ the gain depends quadratically on the length shown in panel (c) dashed lines and solid blue line. For the flux tuning of $\Phi/\Phi_0=0.5$ exponential gain is impossible at all frequencies. \label{fig3}}
\end{figure}
 
Fig \ref{fig3}(a) shows numerical simulations of the signal gain as a function of signal frequency for the proposed TWPA with a transmission line length of $600a$. For a magnetic flux tuning of $\Phi/\Phi_0 =0.45$  and pump power $-76 \text{ dBm}$, Fig. \ref{fig3}(a) (red line) there are two regions $\omega_s/(2\pi)=3.6 \text{ and } 9.4 \text{ GHz}$  where perfect phase matching $\kappa=0$ can be achieved, and for comparison the phase mismatch dependence on signal frequency is shown in Fig. \ref{fig3}(b) (red line). For $\kappa\approx0$ and $g$ is real and large, the gain depends exponentially on the TWPA length Fig. \ref{fig3}(c) (solid red line).  When the phase mismatch is the largest at $\omega_s/(2\pi)=6.5\text{ GHz}$, $g$ is small and $\kappa$ is large, the gain depends quadratically on the length of the TWPA as shown in Fig. \ref{fig3}(c) (dashed red line). Under the phase matching conditions there exist regions of exponential gain and quadratic gain depending on signal frequency, the $3 \text{ dB}$ bandwidth of the TWPA $\sim 1.5 \text{ GHz}$ is limited to two regions where $\kappa\approx0$ centered at $\omega_s/(2\pi)=3.9 \text{ and } 9.1 \text{ GHz}$.With magnetic flux tuning value of $\Phi/\Phi_0 =0.5$ and pump power $-70 \text{ dBm, } \kappa$ is large Fig. \ref{fig3}(b) (blue line) and only a quadratic gain dependence is possible. Since the signal gain increases quadratically for all frequencies a relatively flat gain characteristic of the amplifier can be achieved with a signal gain of $23 \text{ dB}$ over a $3 \text{ dB}$ bandwidth greater than $5.4 \text{ GHz}$.
\begin{figure}[h!]
\includegraphics[width=8.5cm]{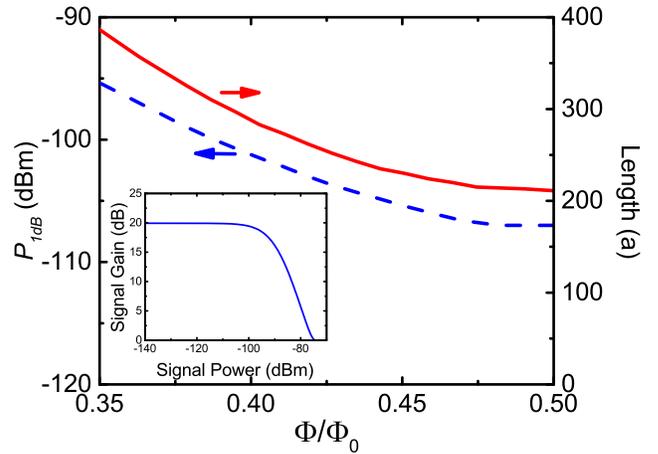}   
\caption{(color online) Calculated $1 \text{ dB}$ compression point and minimum TWPA length as a function of magnetic flux $\Phi/\Phi_0$  to maintain a signal gain of $20\text{ dB}$. As $\Phi/\Phi_0$ varies from $0.35$ to $0.5$ the pump power was varied from $-68 \text{ dBm}$ to $-76 \text{ dBm}$ to maintain phase matching conditions between a signal and pump tone at angular frequencies $\omega_s/(2\pi)= 6 \text{ GHz}$ and $\omega_p/(2\pi)=9 \text{ GHz}$ respectively. The inset shows the signal gain as a function of signal power for $\Phi/\Phi_0 =0.37$ and device length $341a$, the $1 \text{ dB}$ compression point occurs at $-98 \text{ dBm}$. \label{fig4}}
\end{figure}

The saturation power of the TWPA is limited by pump depletion effects, which generally occurs when the signal amplitude becomes comparable to the pump. At such amplitudes the pump is depleted and the gain of the TWPA decreases. In order to evaluate these effects the coupled mode equations (see Appendix B) without the un-depleted pump approximation and taking into account self- and cross-phase modulation are solved to determine the real amplitude and phase mismatch as a function of $z$ along the length of the transmission line. Figure \ref{fig4} (inset) shows how the signal gain decreases with the signal power due to pump depletion effects, for $\Phi/\Phi_0=0.37$ and a TWPA length of $341a$. The signal gain and the phase mismatch depend on magnetic flux through $\alpha_p\propto \gamma k_p^5 |A_{p0}|^2$ and $k_m$. For each  $\Phi/\Phi_0$  the pump power is varied over the range $-76\text{ dBm }(I_{prms}\approx0.1I_{jl0})$ to $-68\text{ dBm }(I_{prms}\approx0.25I_{jl0})$ to maintain phase matching at $\omega_s/(2\pi)=6 \text{ GHz}$  with a $\omega_p/(2\pi)=9 \text{ GHz}$ pump. For each $\Phi/\Phi_0$  the minimum length of the TWPA to achieve a signal gain of $20\text{ dB}$ is shown in Fig. \ref{fig4} (red line). When $\alpha_p$  and $k_m$ decrease with decreasing $\Phi/\Phi_0$  a stronger pump is required to maintain phase matching conditions which results in a larger $P_{1dB}$ up to $-98 \text{ dBm}$, smaller $\gamma$, and a longer transmission line to maintain a signal gain of $20\text{ dB}$. When $\alpha_p$  and $k_m$ increases with $\Phi/\Phi_0$ a smaller pump power is required to achieve phase matching, $ P_{1dB}$ decreases, $\gamma$ increases resulting in a shorter transmission line with a minimum length of $211 a$ to achieve a signal gain of $20\text{ dB}$.
\section{Summary}
In conclusion a TWPA design based on a chain of coupled asymmetric SQUIDs has been presented. The proposed design allows for great flexibility where a magnetic flux can be used to tune the nonlinearity of the transmission line to achieve phase matching conditions in a four-wave mixing process. Numerical simulations have shown that the proposed amplifier can achieve gains of $23\text{ dB}$ with a $3\text{ dB}$ bandwidth of greater than $5.4 \text{ GHz}$ at a center frequency of $6.5 \text{ GHz}$. Under different tuning conditions gains of greater than $20\text{ dB}$ can be achieved with a minimum transmission line length of $211$ unit cells and a saturation power of up to $-98 \text{ dBm}$ with $341$ unit cells. The proposed amplifier is ideally suited for multiplexed readout of quantum bits or kinetic inductance based astronomical detectors.
%
\begin{acknowledgments}
We would like to thank M. Gershenson for helpful discussions. This work was supported in part by Solid State Scientific Corporation (US Army Small Business Technology Transfer Program).
\end{acknowledgments}

\renewcommand{\theequation}{A\arabic{equation}}
\renewcommand{\thefigure}{A\arabic{figure}}
\setcounter{figure}{0}

\begin{appendix}
\section{Analytical Approximation}
In this section we derive an analytical solution to the coupled nonlinear wave equations used to describe the interaction between the signal and pump traveling waves propagating along the transmission line composed of a chain of coupled asymmetric SQUIDs. The energy phase relation of a unit cell Figure 1(a) is
\begin{align}\label{seq1}
\begin{split}
E_J(\varphi) = - 2&E_{jl0} \cos\Big(\frac{\varphi}{2}\Big)- E_{js0}\cos\Big(\varphi-\frac{2\pi\Phi}{\Phi_0}\Big) -  \\ 
&  E_{js0}\cos\Big(\varphi+2\pi\frac{\Phi}{\Phi_0}\Big), 
\end{split}
\end{align}
Expanding $\partial E_J (\varphi)/\partial \varphi|_{\varphi=0}$ gives the approximate current phase relation describing the current $I_n (\varphi)$ flowing through the backbone of unit cell $n$ Eq. (1). Utilizing Kirchhoff's current law $I_{n-1} (\varphi)-C_{gnd} \, d^2 \varphi/dt^2=I_n(\varphi)$ and neglecting the effects of dissipation and assuming a sufficiently long wavelength $\lambda\gg a$ of the signal and pump the following wave equation is derived for a position $z$ along the transmission line
\begin{align}  \label{seq2}
\begin{split}
\frac{a^2}{L}\Big[ \frac{r}{2}+2\cos\Big(\frac{2\pi\Phi}{\Phi_0}\Big)\Big]&\frac{\partial ^2 \varphi}{\partial z^2}+a^2 C_{js}\Big(\frac{r}{2}+2\Big)\frac{\partial ^4 \varphi}{\partial t^2 \partial z^2}- \\ \\
C_{gnd}&\frac{\partial ^2 \varphi}{\partial t^2} -\gamma \frac{\partial}{\partial z}\Big[\Big(\frac{\partial \varphi}{\partial z}\Big)^3\Big]=0,
\end{split}
\end{align}
where $L=\varphi_0/I_{js}\text{, } \varphi_0=\Phi_0/2\pi\text{, } r=I_{jl0}/I_{js0}\text{, and }  \\
\gamma = \big[a^4/(\varphi_0^2 L)\big]\big[(r/48)+(1/3)\cos\big(2\pi\Phi/\Phi_0\big )\big]$. The solution to Eq.(\ref{seq2}) is assumed to be a superposition of a pump, signal, and idler traveling waves propagating along the transmission line of the form
\begin{align}\label{seq3}
\begin{split}
\varphi(z,t)=\frac{1}{2} \Big[ &A_p(z) e^{i(k_p z-\omega_p t)}+A_s(z) e^{i(k_s z-\omega_s t)}+ \\
& A_i(z) e^{i(k_i z-\omega_i t)}+\text{c.c}\Big],
\end{split}
\end{align}
where c.c. denotes complex conjugate, $A_m$ is the complex amplitudes, $k_m$ is the wave vectors, and $\omega_m$ is the angular frequencies of the pump, signal, and idler $(m=\{p,s,i\})$. A degenerate four-wave mixing process is considered under the following frequency matching condition $\omega_s+\omega_i=2\omega_p$. Eq.(\ref{seq3}) is substituted into Eq.(\ref{seq2}) and assuming a slowly varying envelope of the propagating waves where $|\partial^2 A_m/\partial z^2 |\ll|k_m \partial A_m/\partial z| \text{ and } |\partial A_m/\partial z|\ll|k_m A_m |$ and a uniform transmission line where $C_{gnd}$, $C_{js} \text{ and } k_m$ are constant, a set of coupled mode equations which describes the propagation of the pump, signal, and idler waves along the transmission line is determined:
\begin{align}
& \frac{\partial A_p}{\partial z}- i\alpha_p A_p  =0, 	\label{seq4}\\									
& \frac{\partial A_s}{\partial z}-i\alpha_s A_s-\frac{i3\gamma k_p^2 k_i k_s (2k_p-k_i ) A_i^* A_p^2}{8\omega_s^2 C_{gnd}} e^{-i\Delta kz}=0,	\label{seq5} \\
& \frac{\partial A_i}{\partial z}-i\alpha_i A_i-\frac{i3\gamma k_p^2 k_s k_i (2k_p-k_s ) A_s^* A_p^2}{8\omega_i^2 C_{gnd}} e^{-i\Delta kz}=0,	\label{seq6}
\end{align}
where a large pump amplitude relative to the signal and idler amplitudes was assumed, decoupling the pump, and the quadratic terms in $A_{s,i}$ were neglected, $\Delta k=k_s+k_i-2k_p$ is the phase mismatch due to linear dispersion, and $\alpha_m$ is the self-phase modulation per unit length $a$:
\begin{align} \label{seq7}
\begin{split}
& \alpha_s=\frac{3\gamma k_s^3 k_p^2 |A_{p0} |^2}{4C_{gnd} \omega_s^2}, \\	\\						
& \alpha_i=\frac{3\gamma k_i^3 k_p^2 |A_{p0} |^2}{4C_{gnd} \omega_i^2},	 \\	\\						
& \alpha_p=\frac{3\gamma k_p^5 |A_{p0} |^2}{8C_{gnd} \omega_p^2},  
\end{split}
\end{align}
where $|A_{p0}|$ is the initial pump amplitude. The linear dispersion relation for this transmission line is
\begin{align}
k_m=\frac{\omega_m \sqrt{LC_{gnd}}}{a\sqrt{\big[\frac{r}{2}+2\cos\big(2\pi\frac{\Phi}{\Phi_0 }\big)\big]-\omega_m^2 LC_{js}\big (\frac{r}{2}+2\big)}}.\label{seq8}
\end{align}		
Assuming an un-depleted pump amplitude and the following substitutions $A_p (z)=A_{p0} e^{i\alpha_p z}$ solution to Eq. (\ref{seq4}), $A_s (z)=a_s (z) e^{i\alpha_s z} \text{, and } A_i (z)=a_i (z) e^{i\alpha_i z}$ into (\ref{seq5}) and (\ref{seq6}) to obtain:
\begin{align}
& \frac{\partial a_s}{\partial z}-\frac{i3\gamma k_p^2 k_i k_s (2k_p-k_i ) a_i^* |A_{p0} |^2}{8\omega_s^2 C_{gnd}} e^{i\kappa z}=0,	\label{seq9} \\   \nonumber \\
& \frac{\partial a_i}{\partial z}-\frac{i3\gamma k_p^2 k_s k_i (2k_p-k_s ) a_s^* |A_{p0}|^2}{8\omega_i^2 C_{gnd}} e^{i\kappa z}=0, \label{seq10}
\end{align}
where $\kappa=-\Delta k+2\alpha_p-\alpha_s-\alpha_i$ is the total phase mismatch. Equations (\ref{seq9}) and (\ref{seq10}) are similar to well established fiber parametric amplifier theory and have the following solution to describe the amplitude of the signal along the length of the transmission line assuming zero initial idler amplitude:
\begin{align}
a_s (z)=a_{s0} \Big[\cosh (gz)- \frac{i\kappa}{2g} \sinh(gz)\Big] e^{i\kappa z/2} \label{seq11}
\end{align}
A similar solution to (\ref{seq11}) exists for the idler amplitude. The exponential gain factor is
\begin{align}
g=\sqrt{\bigg(\frac{k_s^2 k_i^2 (2k_p-k_s )(2k_p-k_i ) \omega _p^4}{k_p^6 \omega _i^2 \omega _s^2}\bigg) \alpha _p^2-\bigg(\frac{\kappa}{2}\bigg)^2 }. \label{seq12}
\end{align}

\noindent The signal power gain can be determined from Eq. \ref{seq11}

\begin{align}
G_s=\Big|\cosh(gz)-\frac{i\kappa}{2g}\sinh(gz)\Big|^2.		\label{seq13}
\end{align}
For the proposed chain of asymmetric SQUIDs the phase mismatch due to linear dispersion is always real and non-negative for $r>r_0$. Using the frequency matching condition and Eq. \ref{seq8} we show that

\begin{align}
\begin{split}
& \Delta k=c_1 \bigg(\frac{\omega_s}{\sqrt{1-c_2 \omega_s^2}}+\frac{\omega _i}{\sqrt{1-c_2 \omega _i^2}}-\frac{2\omega _p}{\sqrt{1-c_2 \omega _p^2}}\bigg), \nonumber \\
& \text{where } c_1=\frac{\sqrt{LC_{gnd}}}{a\sqrt{\frac{r}{2}+2\cos(2\pi \frac{\Phi}{\Phi_0})}} \text{ and } \\  \\
& c_2=\frac{LC_{js} \big(\frac{r}{2}+2\big)}{\big(\frac{r}{2}+2\cos\big(2\pi \frac{\Phi}{\Phi_0 }\big)\big)} \text{, }\\ \\
&\Delta k \approx 3 c_1 c_2 \Delta \omega ^2 \omega _p \geq 0, \nonumber
\end{split}
\end{align}
where $\Delta \omega \equiv \omega _s - \omega _p=\omega _p - \omega _i$ from the frequency matching condition. We also show that the sign of the first term in Eq.(6) is only dependent on $\alpha _p \text{ for } r > r_0$,
\begin{align}
& \frac{k_m^3}{\omega _m^2}=\frac{c_1 \omega _m}{\big(1-c_2 \omega _m \big)^\frac{3}{2}} \approx c_1 \omega _m \Big(1+\frac{3c_2 \omega _m^2}{2}\Big) \nonumber \\ \\
& \frac{k_s^3}{\omega _s^2}+\frac{k_i^3}{\omega _i^2}-\frac{k_p^3}{\omega _p^2} \approx c_1 \omega _p \Big(1+9c_2 \Delta \omega ^2+\frac{3c_2 \omega _p^2}{2}\Big)>0. \nonumber
\end{align}

\section{Numerical Analysis}
Below the solutions to the coupled mode equations which govern the degenerate four wave mixing process between the signal, idler, and pump tones and pump depletion effects are presented. Plugging Eq. \ref{seq3} into Eq. \ref{seq2} and making the slowly varying envelope approximation the following coupled mode equations are derived:
\begin{align}
\begin{split}
\frac{\partial A_p}{\partial z} - i\frac{3\gamma k_p^2}{8\omega _p^2 C_{gnd}}\Big[k_p \Big(k_p^2 |A_p |^2+ 2k_s^2 |A_s |^2+ 2k_i^2 |A_i |^2 &\Big) A_p+\\ 
 2k_s k_i \Big(k_s+k_i-k_p \Big) A_p^* A_s A_i e^{i\Delta kz} &\Big]=0, \nonumber \\ \\
\frac{\partial A_s}{\partial z} - i \frac{3\gamma k_s^2}{8\omega _s^2 C_{gnd}}\Big[k_s \Big(2k_p^2 |A_p |^2+k_s^2 |A_s |^2+
2k_i^2 |A_i |^2 &\Big) A_s + \nonumber\\ 
k_p^2 k_i  \frac{(2k_p-k_i )}{k_s}  A_i^* A_p^2 e^{-i\Delta kz} & \Big]=0, \nonumber \\ \\
\frac{\partial A_i}{\partial z} - i \frac{3\gamma k_i^2}{8\omega _i^2 C_{gnd}} \Big[k_i \Big(2k_p^2 |A_p |^2+2k_s^2 |A_s |^2+  k_i^2 |A_i |^2 &\Big) A_i +\nonumber \\
k_p^2 k_s \frac{(2k_p-k_s )}{k_i}  A_s^* A_p^2 e^{-i\Delta kz}&\Big]=0. \nonumber
\end{split}
\end{align}
The coupled complex differential equations are solved by converting the complex amplitudes to $A_m=B_m (z) e^{i\theta _m (z)}$ and finding the solutions for the real amplitudes $B_p (z)$, $B_s (z)$, $B_i (z)$, and the total phase mismatch $\theta _s (z)+\theta _i (z)-2\theta _p (z)+\Delta k$ using the Runge-Kutta method.

\section{Parameter Variations}
In this section we determine how variations in Josephson junction parameters and tuning magnetic flux gradients effect the operation of the proposed TWPA. Both types of variations manifest themselves as a position dependent alteration to both the linear and nonlinear dispersion from unit cell to unit cell which ultimately effects the phase matching and the signal gain in the TWPA. Since the transmission line of the TWPA is made up of coupled asymmetric SQUIDs variations in one unit cell through mutual coupling, effects the superconducting phase distribution in neighboring unit cells.

We model the variations in Josephson junction parameters as a normal distribution with mean critical currents $I_{js0}$ and $I_{jl0}$ for small and large Josephson junctions respectively, and a relative standard deviation $\sigma_{JJ}$ which applies to both small and large junctions. The proposed TWPA was modeled by minimizing the total Josephson energy of a 200 unit cell chain by adjusting the phases on all the large junctions. A gauge was chosen such that the external magnetic flux would induce phases $2\pi\Phi/\Phi_0$ on the small junctions. The constraint of the minimization procedure was that the sum of the phase drops on all of the large junctions would be equal to the total phase across the chain. From the minimization a position dependent $\Delta k_m$ and $\Delta \gamma$ from unit cell to unit cell was determined, vectorized, and introduced into the numerical analysis of the coupled mode equations.

 \begin{figure}[h!]
 \includegraphics[width=8.5cm]{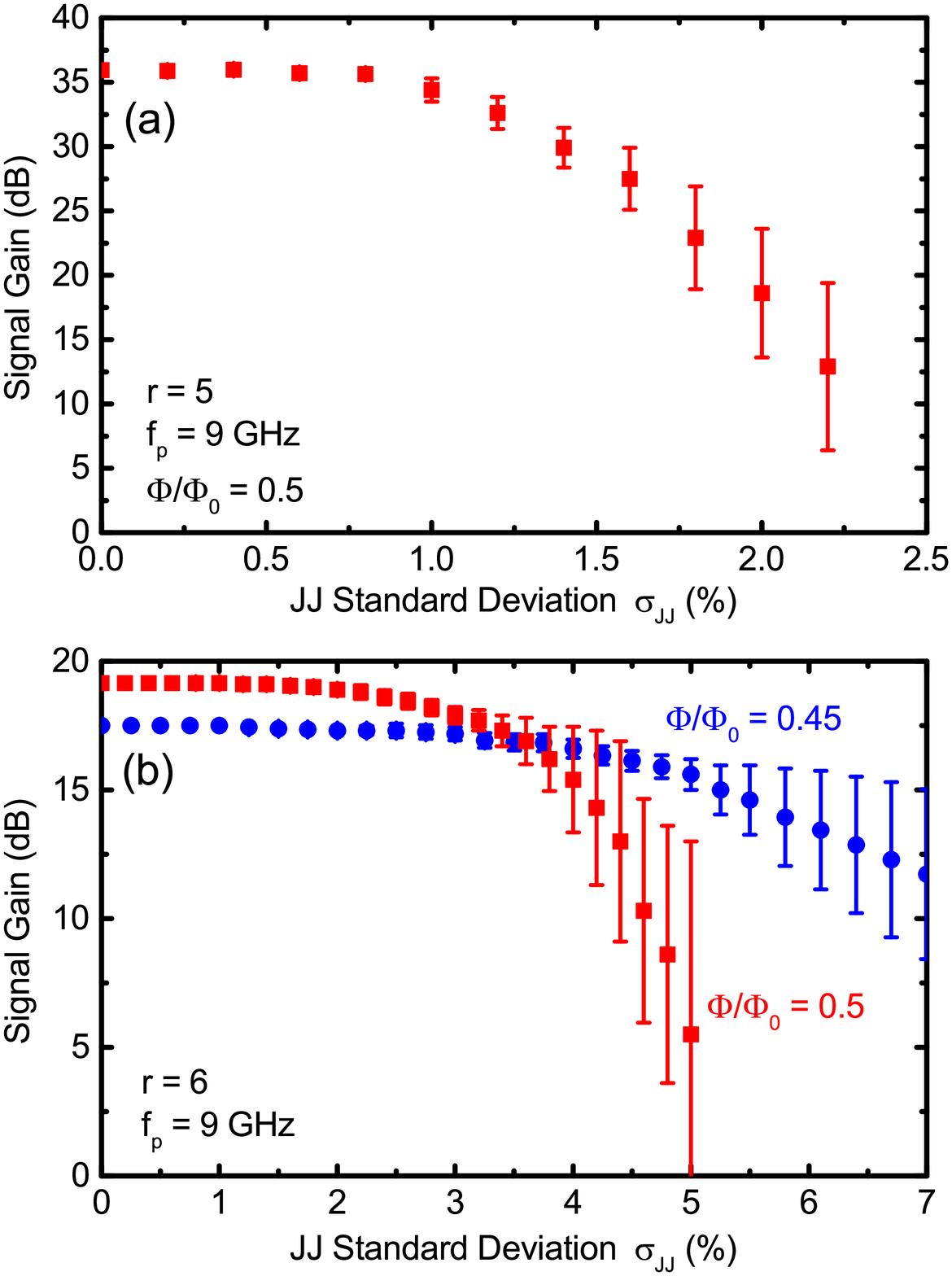}   
 \caption{Simulated signal gain as a function of the relative standard deviation of the critical currents of both the small and large Josephson junctions which make up the TWPA. Simulations were performed at $\Phi/\Phi_0 = 0.45$ (blue circles) and $\Phi/\Phi_0  = 0.5$ (red squares), $\omega_p/(2\pi)=9 \text{ GHz, } \omega_s/(2\pi)=6 \text{ GHz,}$ TWPA length $200\text{a}$, and $r = 5 \text{ and } r = 6$ for Panel (a) and (b) respectively. The error bars represent the variation in signal gain due to the stochastic solution of the coupled mode equations to a normal distribution of junction parameters along the length of the TWPA.\label{figA1}}
 \end{figure}
 
  \begin{figure}[h!]
 \includegraphics[width=8.5cm]{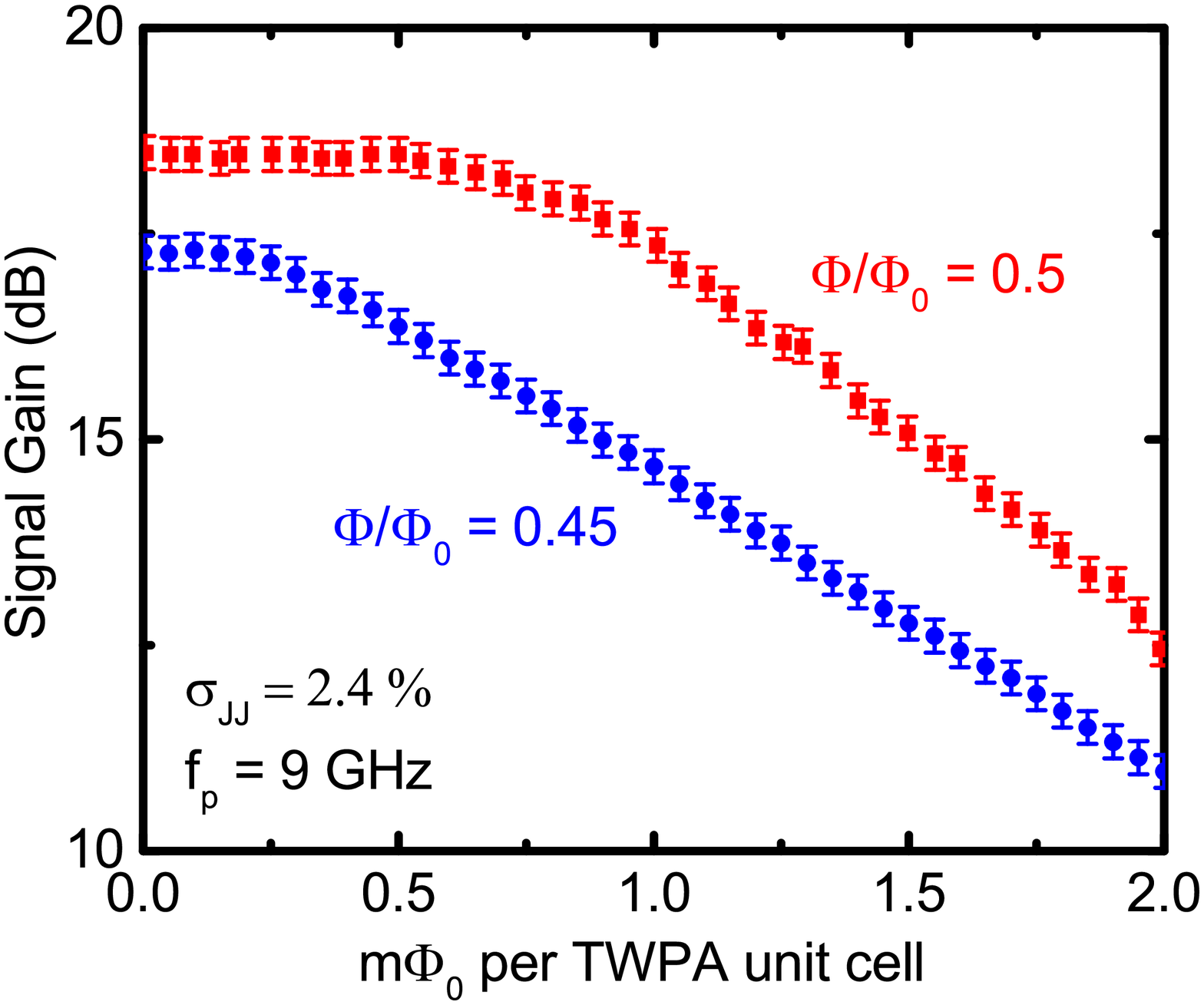}   
 \caption{Simulated signal gain as a function of magnetic flux gradient per unit cell across a TWPA composed of 200 unit cells, for magnetic flux tunings of $\Phi/\Phi_0   = 0.45$ (blue circles) and $\Phi/\Phi_0  = 0.5$ (red squares). Error bars represent the stochastic solutions of the coupled mode equations due to a normal distribution in Josephson junction parameters with $\sigma_{JJ} = 2.4 \%$. \label{figA2}}
 \end{figure}

 \begin{figure}[h!]
 \includegraphics[width=8.5cm]{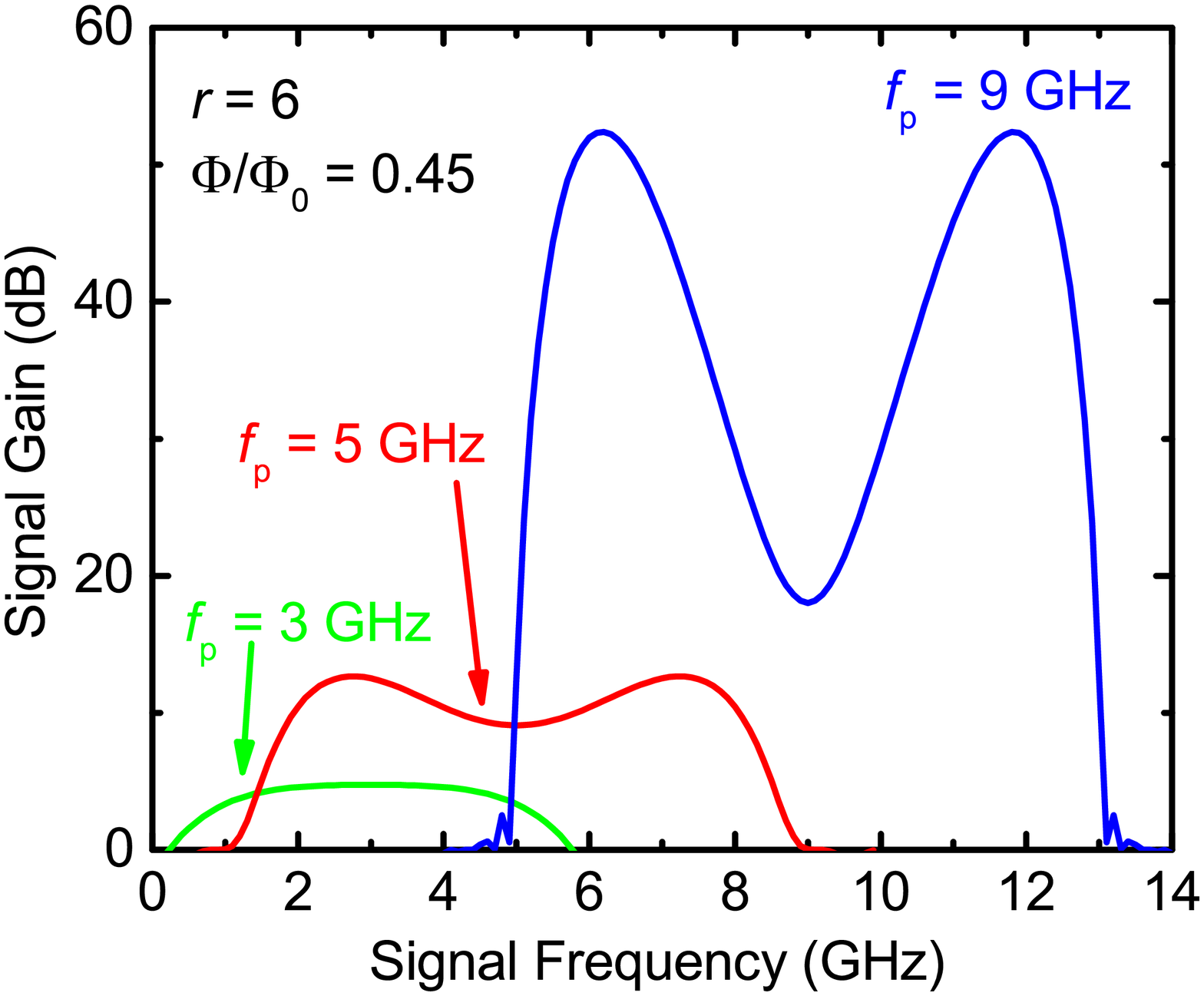}   
 \caption{Numerical simulation of the signal gain as a function of signal frequency for pump frequencies $\omega_p/(2\pi)=3 \text{ GHz (green line), } 5 \text{ GHz (red line), and } 9 \text{ GHz (blue line)}$ respectively, pump power of $-76 \text{ dBm, }r = 6$,  TWPA length of $600a$ and $\Phi/\Phi_0 = 0.45$. \\  \label{figA3}}
 \end{figure}

Numerical simulations of the TWPA which take into account a normal distribution in Josephson junction parameters were performed on the shortest $200a$ TWPA with a $\omega_p/(2\pi)=9 \text{ GHz}$ and pump power of $-76 \text{ dBm}$. Shown in FIG. \ref{figA1}(a) and FIG. \ref{figA1}(b) is the signal gain at $\omega_s/(2\pi)=6 \text{ GHz} )$ where $\kappa\approx 0 $ as a function of the relative standard deviation $\sigma_{JJ}$ for different $r$ and $\Phi$. For each data point 50 numerical simulations were performed and the error bars represent the spread in signal gain due to the stochastic nature of the solution to the coupled mode equations to a normal distribution in junction parameters. FIG. \ref{figA1}(a) shows for $r = 5 \text{ and } \Phi/\Phi_0 = 0.5$  the tolerance of the proposed TWPA to a variation in junction parameters is limited to $\sigma_{JJ} = 1\%$, where the signal gain drops by $1 \text{ dB}$. FIG. \ref{figA1}(b) shows for $r = 6 \text{ and } \Phi/\Phi_0 = 0.5$, the tolerance increases to $\sigma_{JJ} = 3 \%$ which can be realized with present day fabrication technology where the on-chip variation in junction parameters follows a normal distribution and $\sigma = 2.4-3.5 \%$ \cite{BellPRL12,BellPRL14,BellPRB12,BellArXiv15,PopJVSTB12}. When the magnetic flux is tuned to $\Phi/\Phi_0 = 0.45$ the nonlinearity is decreased, however, the tolerated $\sigma_{JJ}$ increases to $4.5 \%$. A trade-off exists between the magnitude of the nonlinearity (for small $r$ or large $\Phi$) and the tolerance of the TWPA to a variation in junction parameters.

Numerical simulations of the signal gain as a function of magnetic flux gradient per unit cell across the TWPA is shown in FIG. \ref{figA2} for tuning magnetic fluxes of $\Phi/\Phi_0 = 0.45$ (blue circles) and $\Phi/\Phi_0 = 0.5$ (red squares). The error bars represent the stochastic nature of the numerical simulations of the coupled mode equations due to a normal distribution in Josephson junction parameters with $\sigma_{JJ} = 2.4 \%$. At $\Phi/\Phi_0 = 0.5$ the TWPA has a 1 dB drop in signal gain for a magnetic flux gradient of $1 \text{ m}\Phi_0$ per unit cell. When the magnetic flux is tuned to $\Phi/\Phi_0 = 0.45$ where the TWPA is more sensitive to magnetic flux variations, the signal gain drops by 1 dB at $0.5 \text{ m}\Phi_0$  per unit cell. In our measurement setup with moderate magnetic shielding we have observed magnetic flux gradients of $24 \text{ } \mu \text{Gauss}/\mu \text{m}$ \cite{BellPRL12,BellPRL14,BellPRB12,BellArXiv15} which is equivalent to $0.1 \text{ m}\Phi_0$  per unit cell of the TWPA.

\section{Different Pumping Conditions}

\noindent Numerical simulations of the signal gain as a function of signal frequency is shown in FIG. \ref{figA3} for alternative pump frequencies $\omega_p/(2\pi)=3 \text{ GHz, } 5 \text{ GHz, and } 9 \text{ GHz}$ to be compared with the results shown in FIG. 3(a). Under phase matching conditions, $\kappa \approx 0$, the maximum in the signal gain depends exponentially on $g$ (see Eq. 5), which has a strong dependence on $\omega_p$. FIG. \ref{figA3} shows that the maximum gain and bandwidth increase with $\omega_p$ and thus shorter transmission lines can be realized as was demonstrated in FIG. 4. For smaller $\omega_p$ the maximum gain and bandwidth decreases and thus longer transmission lines are required even under optimal tuning conditions for such pump frequencies.
\end{appendix}

\bibliography{twpa_v3ref}
\end{document}